\documentstyle[epsfig]{mn}
\input epsf
\title[The inner bulge]
{An excess of very bright stars in the inner bulge}

\author[L\'opez-Corredoira et al.]{M. L\'opez-Corredoira
\thanks{Electronic mail: martinlc@ll.iac.es.}$^1$, F.
Garz\'on$^{1,2}$, P. L. Hammersley$^1$\\
$^{1}$ Instituto de Astrof\'\i sica de Canarias, E-38200  La 
Laguna, Tenerife, Spain\\
$^{2}$ Departamento de Astrof\'\i sica. Universidad de La Laguna, 
E-38204 La Laguna, Tenerife, Spain}

\date{Accepted xxxx.
      Received xxxx;
      in original form xxxx}

\begin{document}

\maketitle

\begin{abstract}
From an analysis of the stars remaining in central regions of the
Galaxy after subtracting those belonging to the disc and the bulge, we
deduce that the inner bulge must have an extra young population
with respect the rest of the bulge.  
It is shown that there is a higher ratio of very bright stars
in the central bulge than than there is in the outer bulge. This
is interpreted as being an additional young component due to the
presence of star formation regions near the Galactic Centre which is
absent in the outer bulge.
\end{abstract}


\section{Introduction}

In previous papers
(L\'opez-Corredoira et al. 1997b; L\'opez--Corredoira et al. 2000 
(hereafter L00)) 
we have analysed the luminosity function and density distribution of 
the Galactic bulge by inverting  $K$-band star 
counts from the Two Micron Galactic Survey 
in a number of off-plane regions ($2^\circ <|b|<10^\circ $). 
Assuming a non-variable luminosity function within the bulge,
we derived the top end of the $K$-band luminosity function and 
the stellar density function, whose morphology was fitted to 
triaxial ellipsoids. The luminosity function shows a sharp decrease brighter 
than $M_K=-8.0$ when compared with that assumed for the disc population.

In this paper, we provide an analysis of the central plane regions,
i.e. including $|b|<2^\circ $, in order to show that the innermost regions of
the bulge have different characteristics from those of the outer bulge.
The study of the inner bulge cannot be carried out by means of inversion
techniques because the extinction is very patchy there, and we cannot carry
out an inversion of the whole available data of the bulge because
the hypothesis of constant luminosity function, which is a good approach
for the outer bulge, is not for the bulge as a whole.

\section{TMGS data}

\begin{figure}
\begin{center}
\mbox{\epsfig{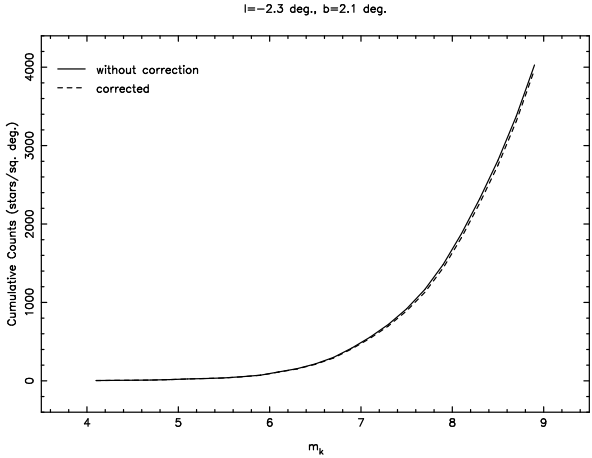}}
\end{center}
\caption{Comparison of cumulated star counts without confusion correction
and those corrected according to the method explained in 
L\'opez-Corredoira et al. (1997a) with a linear extrapolation
of the differential star counts over magnitude 9.4.} 
\label{Fig:corrcrowd}
\end{figure}

$K$-band star counts were obtained from the Two Micron Galactic Survey (TMGS,  
Garz\'on et al. 1993, 1996; Hammersley et al. 1999), 
which has covered more than 350 square degrees of sky and has detected some 
700000 stars in or near the Galactic plane. 
The survey is made up of constant  declination strips which cross the
Galactic plane in the areas $-5^\circ <l<35^\circ $, $|b|\le 15^\circ$ and 
$35^\circ <l<180^\circ $, $|b|\le 5^\circ $.
It is complete between the limits $m_K=4.0$ and 
$m_K\approx 9.2$, except for the regions near the Galactic centre 
where confusion reduced the faint limit by about half a magnitude (although the
detection limit of the survey is fainter than 10 mag). 
The method described in L\'opez-Corredoira et al. (1997a) show 
that confusion has a negligible effect for magnitudes brighter than these 
limits. Figure \ref{Fig:corrcrowd} shows both the 
corrected and uncorrected counts at $l=-2.3^\circ $, $b=2.1^\circ $. 
As can be seen in the figure, the counts are nearly the same with or 
without correction.
That confusion is not significant for the areas chosen can also be
seen in the figures in Hammersley et al. (1999) when the TMGS counts are 
compared with a model. Taking into account that the correction is based 
on an extrapolation and the changes are minor when compared to the other 
sources of error, it is preferable to avoid any correction and use the 
original counts.

Overcrowding effects are not important here since the confusion is produced
when two or more stars are placed in an area rather less than $S=15"\times 15"=
1.7\times 10^{-5}$ deg$^2$ 
\footnote{The detector size of the TMGS is 15". However, we can distinguish
stars with separations less than that (see Garzón et al. 1993) in right
ascension (around 4" or 5") because we sweep the sky in this direction
and the peaks of closer stars can be distinguished.}. 
If we assume a Poissonian distribution in a field of density
$n=4000$ star/deg (the maximum observed density up to magnitude 8.5), 
the probability of observing two stars as one is less than 
$P=S^2n^2/2=4.6\times 10^{-3}$ (L\'opez-Corredoira et al. 1997a).
Therefore, the artifacts of image crowding which generate possible excess 
of star counts (DePoy et al. 1993) cannot be responsible of the
remnants to be studied along this paper.
DePoy et al. (1993) are arguing that clustering of stars will increase 
this effect, however in L\'opez-Corredoira et al. (1998)
we looked for clustering in the TMGS and found little in the bulge.

The constant central declination strip ($\delta=-29^\circ 44'$,
cutting the plane at $l=-0.9^\circ $) with a width of
$\Delta \delta=2.51^\circ $ (hence including the Galactic centre)
within $|b|<10^\circ $ is used in this paper 
to provide information on the innermost region
of the Galaxy. 

\begin{figure}
\begin{center}
\mbox{\epsfig{file=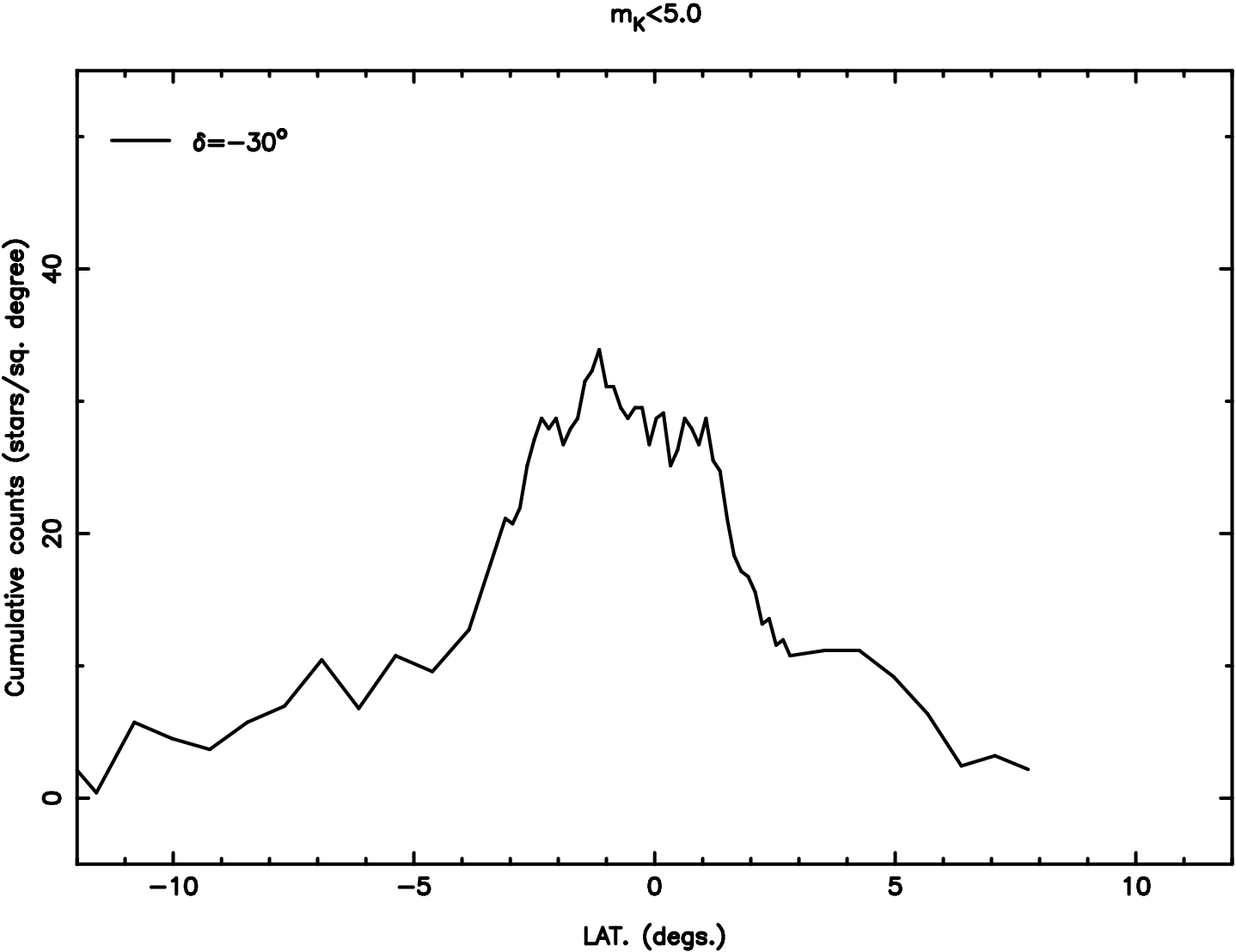,height=8cm,angle=-90}}

Fig. \ref{Fig:cuentas} a)
\end{center}
\end{figure}
\begin{figure}
\begin{center}
\mbox{\epsfig{file=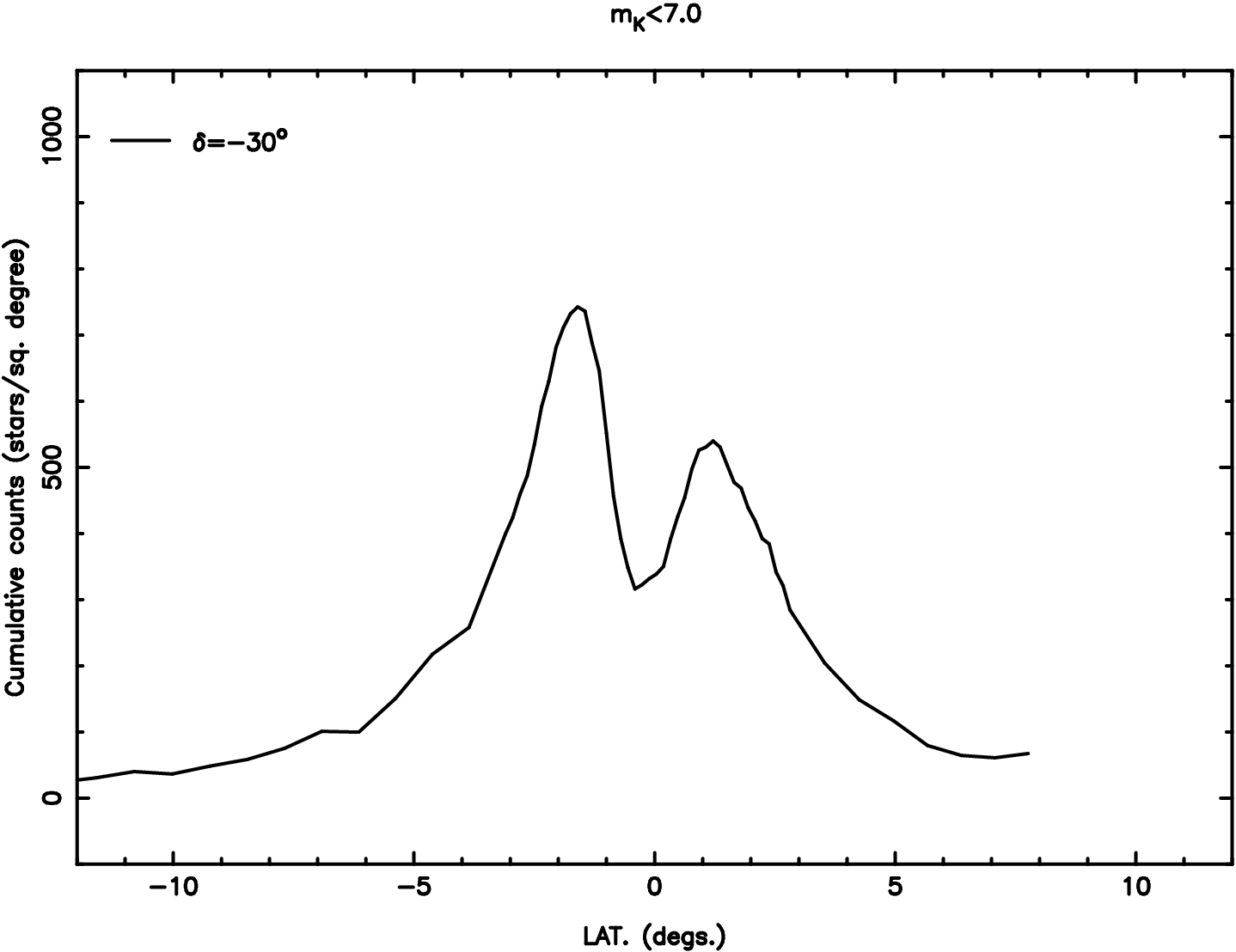,height=8cm,angle=-90}}

Fig. \ref{Fig:cuentas} b)
\end{center}
\end{figure}
\begin{figure}
\begin{center}
\mbox{\epsfig{file=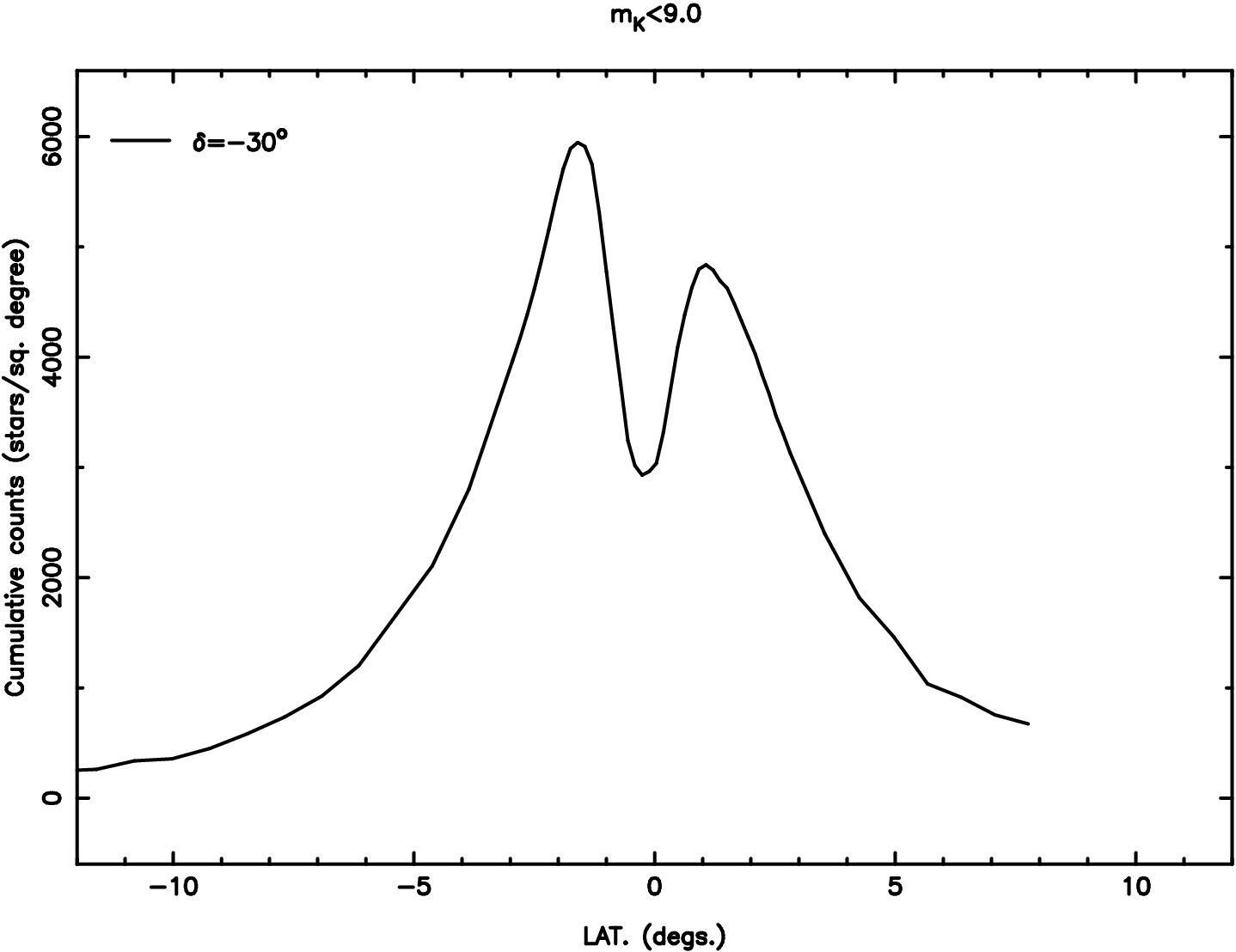,height=8cm,angle=-90}} 

Fig. \ref{Fig:cuentas} c)
\end{center}
\caption{Cumulative counts versus Galactic latitude 
along the strip with constant
declination $\delta =-30^\circ $, 
which cuts the plane at $l=-1^\circ$.
a) Up to 5th magnitude; b) up to 7th magnitude; c) up to 9th
magnitude.}
\label{Fig:cuentas}
\end{figure}

\subsection{Disc and bulge subtraction}

In order to examine any remaining components in the inner Galaxy, first
modelled disc and bulge stellar components are subtracted 
from the TMGS star counts.

The model of the disc, although coded by us, is based on 
the Wainscoat et al. (1992) SKY code. This model
provides a good fit to the TMGS counts in the region where the disc 
dominates (Cohen 1994b; Hammersley et al. 1999; L00). 
The Wainscoat et al. (1992) model was revised by 
Cohen (1994a) but this does not significantly 
alter the form of the disc in the areas of interest here, and
 it is reasonable to expect that this disc model will 
adequately reflect the disc contribution along the lines of sight 
used in this paper.

A triaxial bulge is from the model by L00. This was obtained from the 
inversion of TMGS star counts after subtracting the disc in central 
off-plane regions. The model applies two approaches to the 
bulge: 1) the fitting of triaxial ellipsoids with constant axial ratios to
the bulge and 2) the introduction of variable major--minor axial ratios
to get a best fit to the off-plane regions. Here we use the variable axial
ratio model, however either model fit leads to nearly the same qualitative 
result.

Extinction is included in the same manner as is  
Wainscoat et al. (1992) model. This  assumes  
an exponential distribution of extinction with the same scale length as the 
old disc, $3.5$ kpc, and a scale height of 100 pc. The extinction is then 
normalized to give $A_K=\frac{da_K}{dr}=0.07$ mag Kpc$^{-1}$ in 
the solar neighbourhood. In fact, this model may lead
to an overestimation of the extinction in the inner disc 
(Hammersley et al. 1999), but by comparing  different regions it will 
be shown that this cannot be the cause of the excess of stars in the
central bulge.

Figure \ref{Fig:rem} shows the residual counts after subtracting  both the 
disc and bulge.
The counts versus Galactic latitude present an ``M''-shape, with centre   
on the plane. Furthermore, these counts 
are mostly between $m_K=5$ and $m_K=7$. Figure \ref{Fig:pgplota}
shows this for one region. Up to $\sim 200$ stars/deg$^2$ 
are present in this range of magnitudes in the residuals, which amount to 
an important portion of the total number (including the disc and bulge), 
less than a thousand stars (see Fig. \ref{Fig:cuentas}).
The residuals up to 9th magnitude are relatively low, around
300 stars from a total of 6000, and is barely above the
Poissonian noise. However, between magnitudes 5 and 7,  the residuals are
too excessive to be considered noise or errors in the bulge or disc subtraction.

\begin{figure}
\begin{center}
\mbox{\epsfig{file=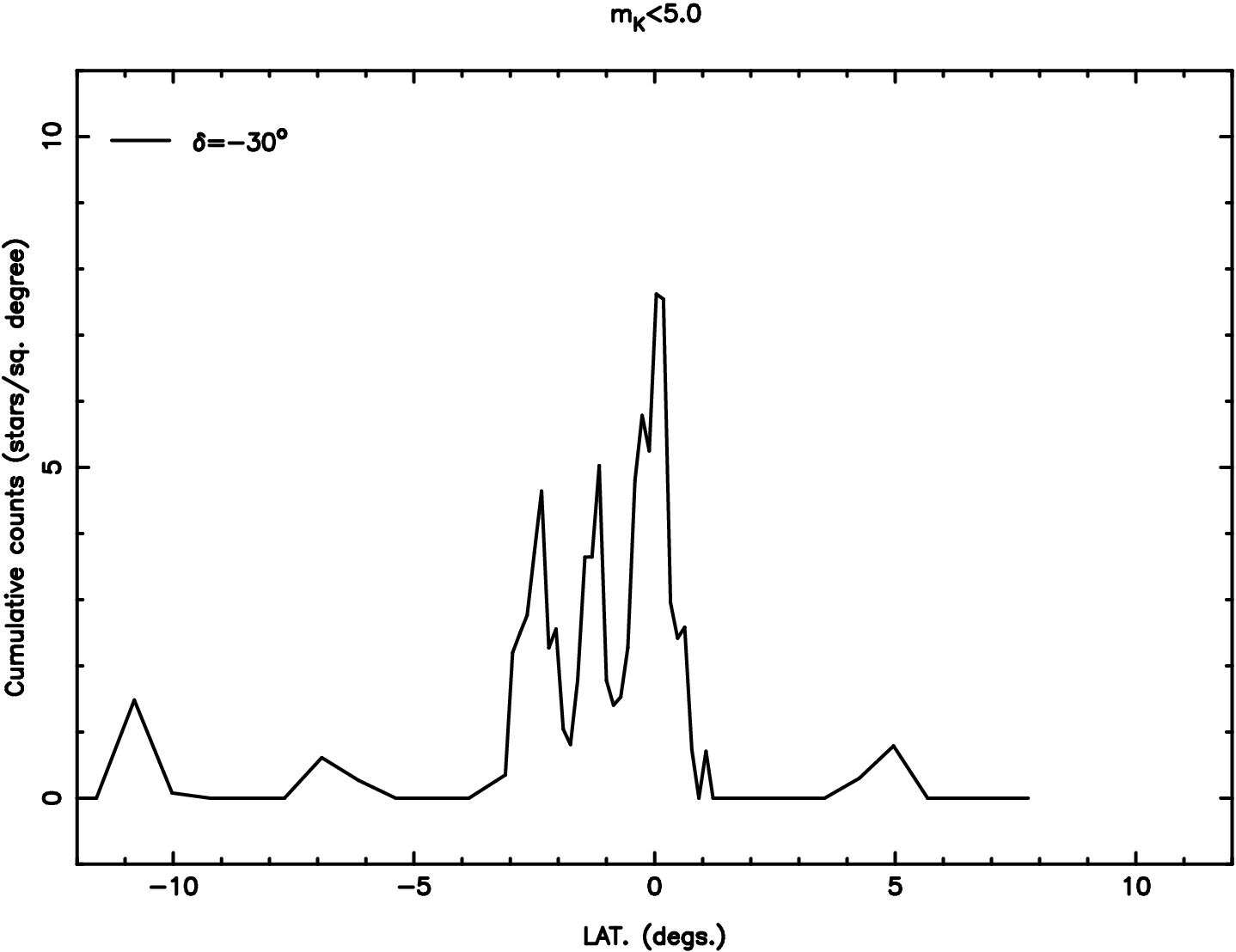,height=8cm,angle=-90}}

Fig. \ref{Fig:rem} a)
\end{center}
\end{figure}
\begin{figure}
\begin{center}
\mbox{\epsfig{file=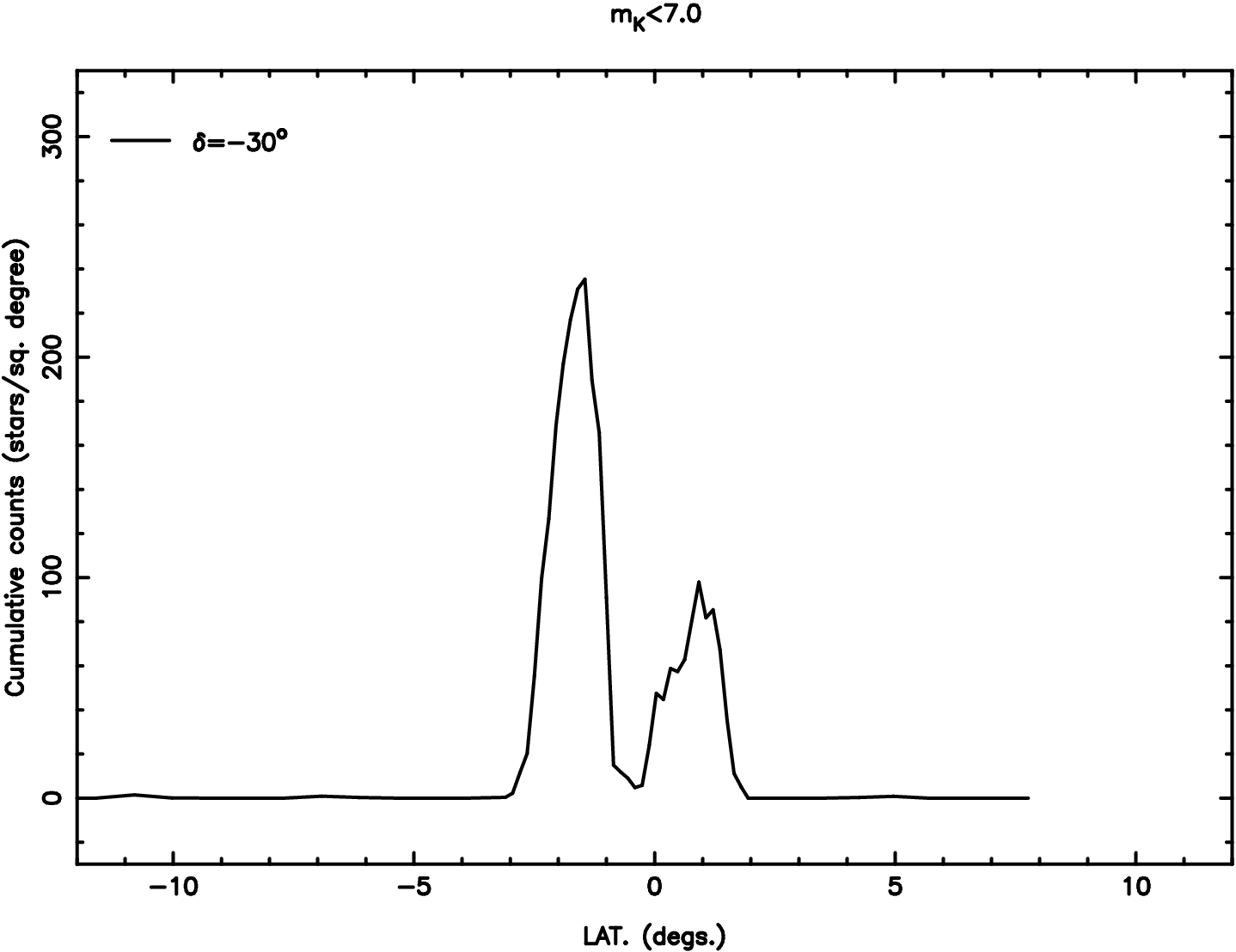,height=8cm,angle=-90}}

Fig. \ref{Fig:rem} b)
\end{center}
\end{figure}
\begin{figure}
\begin{center}
\mbox{\epsfig{file=rem9.ps,height=8cm,angle=-90}} 

Fig. \ref{Fig:rem} c)
\end{center}
\caption{Cumulative counts versus Galactic latitude
as in Fig. \ref{Fig:cuentas} in the strip where the sky was surveyed, 
once subtracted of
Wainscoat et al. (1992) disc model and
L00 bulge model
with variable major--minor axial rate.
a) Up to 5th magnitude; b) up to 7th magnitude; c) up to 9th
magnitude.}
\label{Fig:rem}
\end{figure}

\begin{figure}
\begin{center}
\mbox{\epsfig{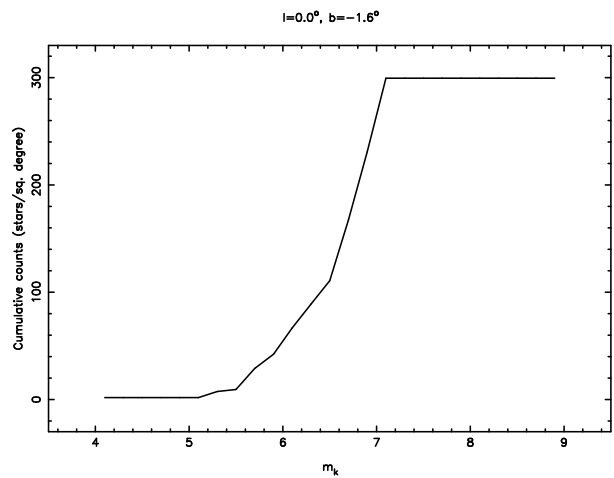}}
\end{center}
\caption{Cumulative counts as in Fig. \ref{Fig:rem}
versus apparent $K$ band magnitude at $l=0.0^\circ $, $b=-1.6^\circ $.}
\label{Fig:pgplota}
\end{figure}

\begin{figure}
\begin{center}
\mbox{\epsfig{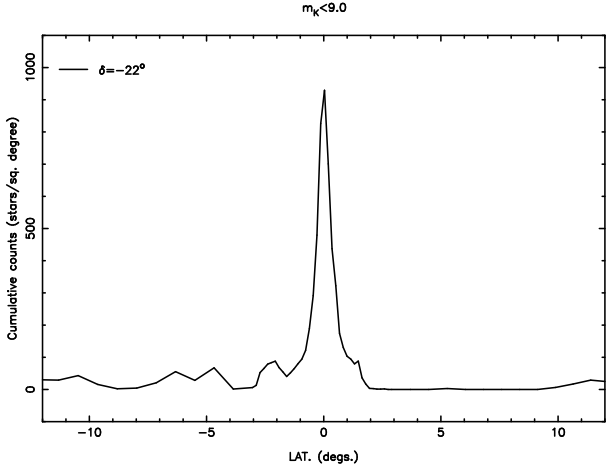}}
\end{center}
\caption{Cumulative counts versus Galactic latitude
along the strip with constant declination $\delta =-22^\circ $, which
cuts the plane at $l=7.5^\circ $, and up to 9th magnitude, once subtracted of
Wainscoat et al. (1992) disc model and
L00 bulge model
with variable major--minor axial rate.}
\label{Fig:res22_9}
\end{figure}

\subsection{An inner bulge and clouds}\
\label{.resinner}

The ``M''-shape is due to clouds near the Galactic centre
(Hammersley et al. 1996, 1999) and this can be clearly seen in the
original data  (Fig. \ref{Fig:cuentas}). The simple extinction model used here 
is not correct in the inner Galaxy including  neither the fall of in 
extinction inside about 3 kpc  nor the dust lanes near the Galactic 
Centre (Hammersley et al. 1999) and hence it clearly shows in the residual 
counts. The strip does not cut the plane exactly in the Galactic centre but 
at $l=-0.9^\circ $ and this leads
to an asymmetry between the peaks of the ``M'', since the positions
with negative Galactic latitudes are closer to the Galactic centre 
than those with positive latitudes. 

The majority of  the remnants are  between magnitudes $K$ 5 and 7 and
there are very few residual stars between  magnitudes 7 and 9.
In fact, most have a magnitude between 6 and 7
(Garz\'on et al. 1993). If these residual stars are at the same distance, 
this would imply that absolute magnitude range is only one or two magnitudes. 
A dispersion in distance and extinction would always increase 
the apparent magnitude range, so the range of absolute magnitudes must 
be limited to a width of one or two.
A Galactic component cannot have such a restricted range of 
absolute magnitudes; hence, here only part of another component is being detected. 
The most probable reason for only detecting a couple of magnitudes
of the LF of this feature is a contrast effect against the old bulge.
L00 shows that the luminosity function of the outer bulge cuts on very 
strongly fainter than about 7 magnitude whereas brighter than this there 
are few stars.  Fainter than about $m_K$=7  
the old bulge component will dominate
every other component making up to 70\% or the sources detected. 
Hence any error in the model for the old bulge leads to  a huge error in any 
residual component fainter than $m_K$=7. However, brighter than $m_K$=7 
there are few old bulge stars and if there were a component in the inner bulge 
extending to brighter magnitudes than the old bulge then it would be most
clearly detected just brighter than $m_K$=7.

The young component in the inner bulge is the best candidate to explain 
these residual stars.

\begin{figure}
\begin{center}
\mbox{\epsfig{file=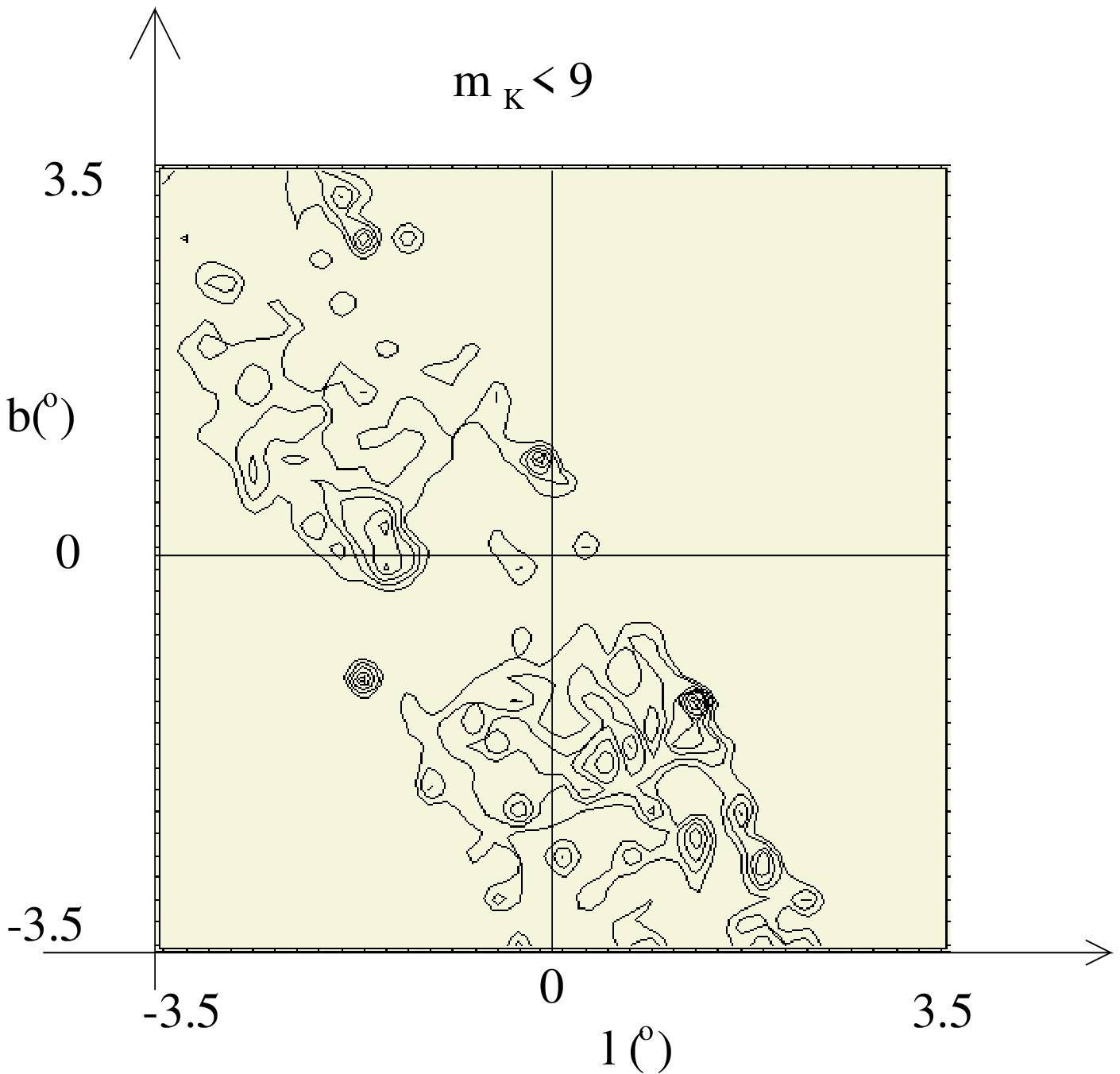,height=7cm}}

Figure \ref{Fig:rescentro} b)
\end{center}
\caption{Contour diagram representing star counts (deg$^{-2}$) for a TMGS strip crossing
the Galactic centre, after subtracted Wainscoat et al. (1992) model and
L00 bulge model
with variable major--minor axial rate.
a) Cumulative counts  to 7th-magnitude. The contours represent steps of
 100 deg$^{-2}$, beginning at zero.
b) Cumulative counts  to 9th-magnitude, the contour steps representing
 200 deg$^{-2}$, beginning at zero.}
\label{Fig:rescentro}
\end{figure}

The young disc (i.e. the spiral arms) cannot be responsible
because the shape of the residual star count would be different, not
an ``M''-shape (Hammersley et al. 1999). Two hundred stars per square
degree up to 7th magnitude (a third of the total number) is too high a
count for the arms , and the ``M''-shape suggests a remote distance for
the sources, beyond the clouds and/or the dust lane. These stars are
located around the Galactic centre, beyond the dust lane. In order to
see this more clearly, Figure \ref{Fig:rescentro} shows  the star
counts with a higher resolution ($0.2^\circ \times 0.2^\circ $) in this
region after subtracting both the model bulge and disc.

\begin{figure}
\begin{center}
\mbox{\epsfig{file=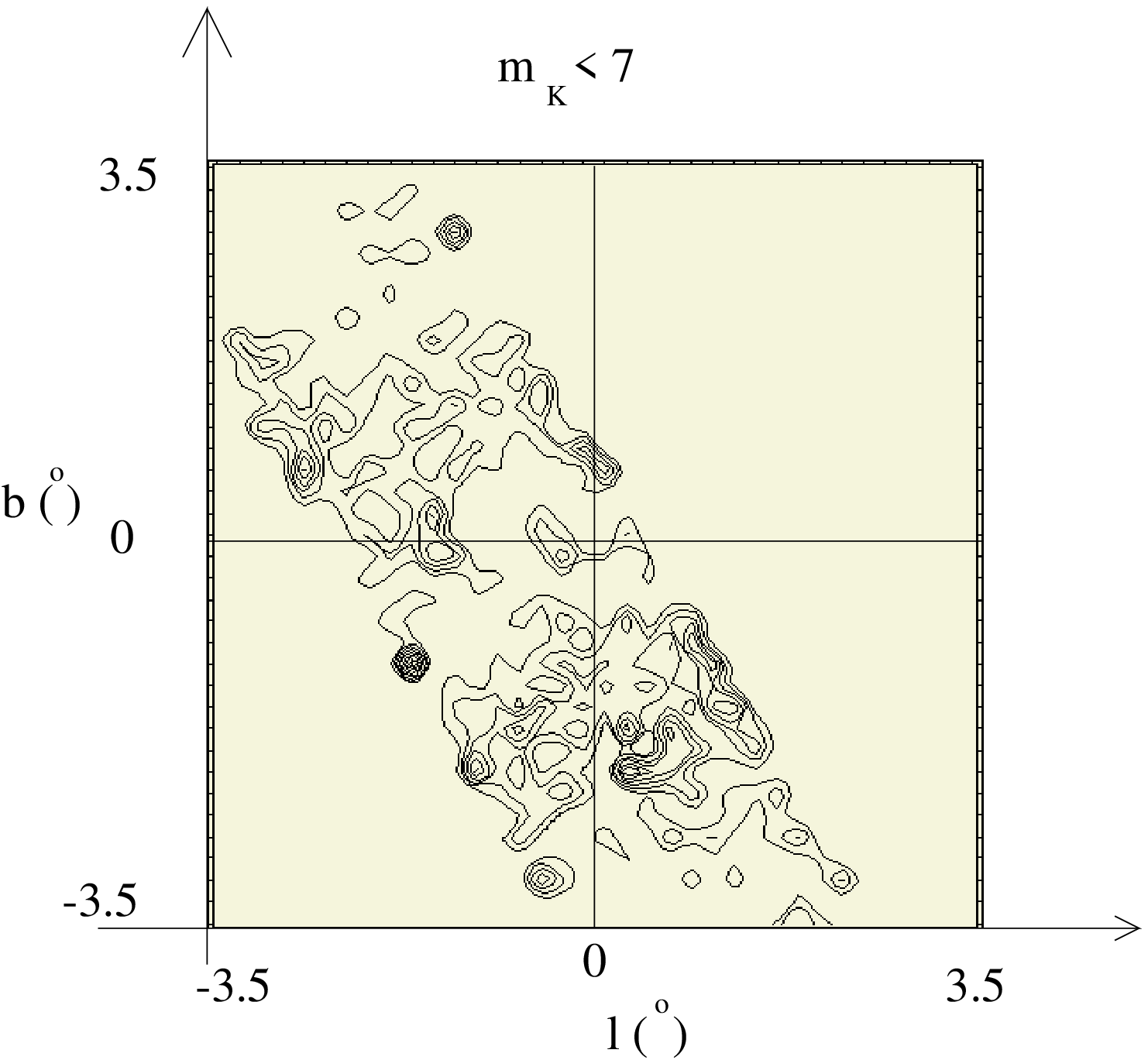,height=7cm}}

Figure \ref{Fig:rescentro} a)
\end{center}
\end{figure}

In Fig. \ref{Fig:rescentro}, both the dust lane and the  increasing
number of counts towards the centre, can be seen.  The residual
distribution is nearly concentric, although the patchiness of the
extinction beaks up  the regularity of the contours. This plot shows
the excess of the ``inner bulge'' with respect the outer bulge. No
accurate morphology of the inner bulge is obtainable because of  this
patchiness and the irregular distribution of young objects in this
component.  This irregularity is also due in part to the Poissonian
error in the star counts in areas of 0.04 deg$^{-2}$; however, this
error disappears when the structure size is larger than the pixel area
(if the excess is correlated with neighbouring pixels, it indicates
that the excess is not random). Nevertheless, the shape can be
distinguished:  from a simple  examination of Fig. \ref{Fig:rescentro}
the inner bulge spreads over $|l|<4^\circ $, $|b|<2^\circ $ and
presumably  has an ellipsoidal morphology.

An overestimation of the extinction in the Wainscoat et al. (1992)
model would also lead to a remnant in the star counts after subtracting
the bulge and disc.  This effect can clearly be seen in the
$l=7.5^\circ $ strip  (see Figure \ref{Fig:res22_9}) and is
discussed in Hammersley et al (1999). However, unlike at $l=-1^\circ $ the
remnant is a narrow spike sitting on the plane and not a broad, `M`
shaped, feature. Furthermore, the extinction would affect the
whole range of magnitudes and not just those with apparent magnitudes
between 6 and 7.  This indicates that the origin for the residual
counts near the GC cannot be extinction.

A young inner bulge population has already been noted 
with the same TMGS data but using a different technique by
Calbet (1993) and Calbet et al. (1995). This population is revealed in
 an excess of very bright stars
($M_K \sim -9$) with respect the rest of the bulge. A gradient of populations
through the whole bulge was considered by L00 as
one possibility to explain the gradient of the major-minor
axial ratio, but this cannot even explain the total number
of stars around the centre, as is shown in this paper; hence, a rather higher
population gradient  must exist in the transition region between the
inner and the outer bulge (at around three degrees from the centre) than that
in the  L00 bulge model.
Later publications have also found the young population with different data
(see Blum et al. 1996a, 1996b; Narayanan et al. 1996; Frogel et al. 1999).
The last-named authors have also pointed out its extension up to one degree away
the Galactic centre, while the stars beyond one degree would be
AGB stars associated with the old bulge population. However, 
we detect an excess of these kinds of stars
up to 2 or possibly 3 degrees away from the Galactic centre, and these belong
to the same young population.

If a depth for the inner bulge of $\sim 1$ kpc is assumed at a distance
of $\sim 8$ kpc, then we deduce an average stellar density of  
$\sim 10^{-5}$ star pc$^{-3}$,  a density which must be added together with
the bulge and disc densities (around one to ten  pc$^{-3}$).
Obviously, the increase in the total number of stars is negligible: 
the increase in mass is more conspicuous but also negligible. 
The only significant contribution is to the flux and the density of very bright
stars.

\subsection{Luminosity function}

An estimate of the luminosity function (useful for qualitative comparison)
may be obtained. Inversion methods for the stellar statistics equation, 
such as that applied in L\'opez--Corredoira 
et al. (1997b, L00), are useful for this purpose.

For each region centred on Galactic coordinates $(l,b)_i$
in the 0.2 degrees resolution plot of Fig. \ref{Fig:rescentro},
where $i$ is the field number, the cumulative  star counts, $N_K$, 
towards the bulge expressed in rad$^{-2}$, follows

\[
N_{K,{\rm bulge}}(m_K)=N_K(m_K)-N_{K,{\rm disc}}(m_K)
\]\begin{equation}
=\int_0^\infty \Phi_{K,{\rm bulge}} (m_K+5-5\log _{10} r -a_K(r))
D_{\, \rm bulge}(r) r^2 dr
\label{sc_acum}
,\end{equation}
where
$\Phi _{K,{\rm bulge}}(M)=\int _{-\infty}^{M}\phi _{K,{\rm bulge}}(M)dM$,
$\phi $ is the normalized luminosity function ($\int _{-\infty}^\infty$
$\phi (M)dM=1$), $D$ is the density, and $a_K$ is the extinction in the 
line of sight. The disc and extinction models are taken from Wainscoat et
al. (1992) model.

With the change of variables
$\rho _K=10^{0.2a_K(r)}r$
and $\Delta _K=D(r)\frac{r^2dr}{\rho _K^2 d\rho _K}$
we transform  equation (\ref{sc_acum})
for counts in the bulge into

\begin{equation}
N_K(m_K)=\int_0^\infty \Phi_K (m_K+5-5\log _{10} \rho _K )
\Delta_K(\rho _K) \rho _K^2 d\rho _K
\label{sc_acum_fic}
,\end{equation}
which is a Fredholm equation of the  first kind.
When  the luminosity function,
$\Phi $, is the unknown, instead of $\Delta $, then 
we can make a new change of variable
$M_K=m_K+5-5\log _{10}\rho _K$ to obtain:

\[
N_{K}(m_K)=200({\rm ln}\ 10)10^{\frac{3m_K}{5}}
\]\begin{equation}\times
\int_{-\infty }^\infty \Delta_K(
10^{\frac{5+m_K-M_K}{5}}) 10^{\frac{-3M_K}{5}}
\Phi_K (M_K)dM_K
\label{lumin_inc}
,\end{equation}
where $\Phi $  is the unknown function and $\Delta _K$ is 
the kernel of this new (first kind) Fredholm integral equation.
The density, $D$, of the bulge is taken from the model with variable 
major--minor axial ratio by L00 to give
$\Delta _K$, and this integral equation is inverted for each position
using Lucy's statistical method (Lucy 1974; L00). 
Obtaining both the bulge density and luminosity function, as in L\'opez--Corredoira
et al. (1997b, L00), is not appropriate here because the extinction and stellar
distribution is too patchy in the plane to make stable inversions.
The inversion of eq. (\ref{lumin_inc}) is more stable than
the inversion of (\ref{sc_acum_fic}), allowing
the luminosity function to be obtained, because
the density distribution is sharply peaked and so the kernel in
(\ref{lumin_inc}) behaves almost as a Dirac delta function: the shape 
of the density distribution does not significantly affect the shape
of the luminosity function.

\begin{figure}
\begin{center}
\mbox{\epsfig{file=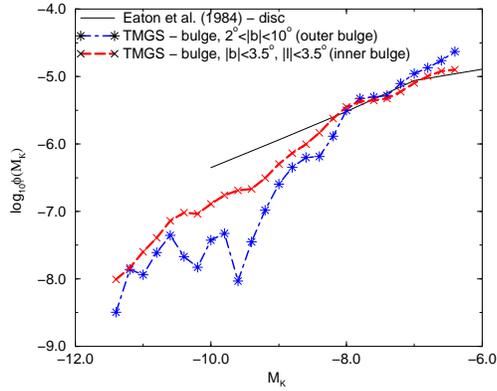,height=6cm}}
\end{center}
\caption{$K$-band luminosity function for disc, inner bulge and outer bulge.}
\label{Fig:lum_inner}
\end{figure}

The average of the luminosity function for all the regions within the
$\delta =-30^\circ $  strip at $|b|<3.5^\circ $, $|l|<3.5^\circ $ is shown in Fig.
\ref{Fig:lum_inner}. 
The outcome is clear: an abundance of
stars with $M_K$ between $-10$ and $-8$
 is observed to be greater than that in the
outer bulge, but less than the abundance in the disc (Eaton et al. 1984). An intermediate abundance between
the young disc---with a high number of very bright giants and supergiants---and the
old disc---with a deficit of these bright stars---is observed from the plot. 
The average includes all regions plotted in Fig.
\ref{Fig:rescentro}, including those within the clouds and/or dust lane
area, so it is slightly underestimated. This suggests the existence of a star formation region in the
centre (Lebofsky, Rieke \& Tokemaga 1982), rich in supergiants and 
bright giants, which is anomalous to the rest of the bulge. 

There are spikes with
residuals an order of magnitude above  the average;
for instance, there are in Fig. \ref{Fig:rescentro} b),  to 9th magnitude,
some points with $\sim $1000 star deg$^{-2}$---around ten times higher than the
average ($\sim $100 star deg$^{-2}$). These spikes would have a luminosity
function with very bright stars comparable with that of the disc, since there
is nearly an order of magnitude between both the disc and inner bulge luminosity
functions. The spikes
in Fig. \ref{Fig:rescentro} are probably related to single star
formation regions, although the patchiness of the extinction is a factor
that can distort the shape of these areas.

\section{Conclusions}

It is shown that the inner bulge contains
an important extra population of bright stars, between around $M_K\sim -10$ and 
$M_K\sim -8$, which are not present in the external bulge. This is 
probably due to the existence of extra star formation near the Galactic Center
but is not present in the rest of the bulge. Hence, in the same manner 
as there is a young and old disc, there appears to be a young inner bulge 
providing an extra contribution to the star counts near the Galactic centre.

\subsection*{Acknowledgments}
We thank the anonymous referee for helpful comments that have
improved the content and presentation of this paper.

\end{document}